\documentclass{article}
\usepackage{spconf,amsmath,graphicx}
\usepackage{multirow}
\usepackage{bm}
\usepackage{cite}
\usepackage{xcolor}

\newcommand{\bhline}[1]{\noalign{\hrule height #1}} 

\title{Rethinking Mean Opinion Scores in Speech Quality Assessment: Aggregation through Quantized Distribution Fitting
}
\name{Yuto Kondo, Hirokazu Kameoka, Kou Tanaka, Takuhiro Kaneko\thanks{This
work was supported by JST CREST Grant Number JP-MJCR19A3, Japan.}}
\address{NTT Corporation, Japan}
\begin{document}
\ninept

\maketitle
\begin{abstract}
Speech quality assessment (SQA) aims to evaluate the quality of speech samples without relying on time-consuming listener questionnaires. Recent efforts have focused on training neural-based SQA models to predict the mean opinion score (MOS) of speech samples produced by text-to-speech or voice conversion systems. This paper targets the enhancement of MOS prediction models' performance. We propose a novel score aggregation method to address the limitations of conventional annotations for MOS, which typically involve ratings on a scale from 1 to 5. Our method is based on the hypothesis that annotators internally consider continuous scores and then choose the nearest discrete rating. By modeling this process, we approximate the generative distribution of ratings by quantizing the latent continuous distribution. We then use the peak of this latent distribution, estimated through the loss between the quantized distribution and annotated ratings, as a new representative value instead of MOS. Experimental results demonstrate that substituting MOSNet’s predicted target with this proposed value improves prediction performance.
\end{abstract}
\section{Introduction}
\label{sec:intro}
\vspace*{-0.2cm}
This study addresses the task of speech quality assessment (SQA), which aims to automatically predict the subjective quality of a given speech sample. Many SQA methods have been proposed to compare qualities of speech processing systems, mainly a conventional system and an improved version of it, without resorting to time-consuming questionnaires to listeners. For example, the perceptual evaluation of speech quality (PESQ)~\cite{rix2001perceptual} is a well-known method that quantifies the relative degradation to reference speech in subjective quality of telephone speech. Recently, with the growing interest in text-to-speech (TTS)~\cite{ren2020fastspeech,popov2021grad} and voice conversion (VC)~\cite{qian2019autovc,kaneko2019stargan,popov2021diffusion} research, numerous SQA models have been proposed for the speech samples after TTS and VC, for which reference speech may not be available in principle. For instance, MOSNet~\cite{lo2019mosnet} predicts the mean opinion score (MOS)~\cite{ITU-TP800} regarding speech quality. MOS is a widely used representative value of speech quality, which is calculated as the mean of ratings obtained by absolute category rating (ACR) method~\cite{ITU-TP800}. The common rating choices for obtaining ACR data include the five options: `1: bad’, `2: poor’, `3: fair’, `4: good’, and `5: excellent’. The ACR is one of the most common and familiar annotation formats in daily life, alongside pairwise comparison.

The target of this paper is to enhance the performance of MOS prediction models. As an approach for the target, we previously explored a new score aggregation method instead of MOS. We focused on the drawback that the reliability of MOS can be limited due to incorrect annotation, and proposed a new representative value of speech quality, the $N$-lowest MOS ($N_{\rm low}$-MOS)~\cite{kondo2024selecting}. $N_{\rm low}$-MOS is calculated as the average of not all ratings but {\it the lowest $N$ ratings}. $N_{\rm low}$-MOS was proposed based on the hypothesis that when humans subjectively rate speech, they tend to assign more weight to low-quality speech segments, and the variance in ratings for each speech sample is primarily attributed to accidental inclusion of higher scores due to overlooking or neglecting the poor quality speech segments. 
\begin{figure*}[t]
\vspace*{0cm}
 \centering
 \hspace*{-0.3cm}
\includegraphics[width=1.4\columnwidth]{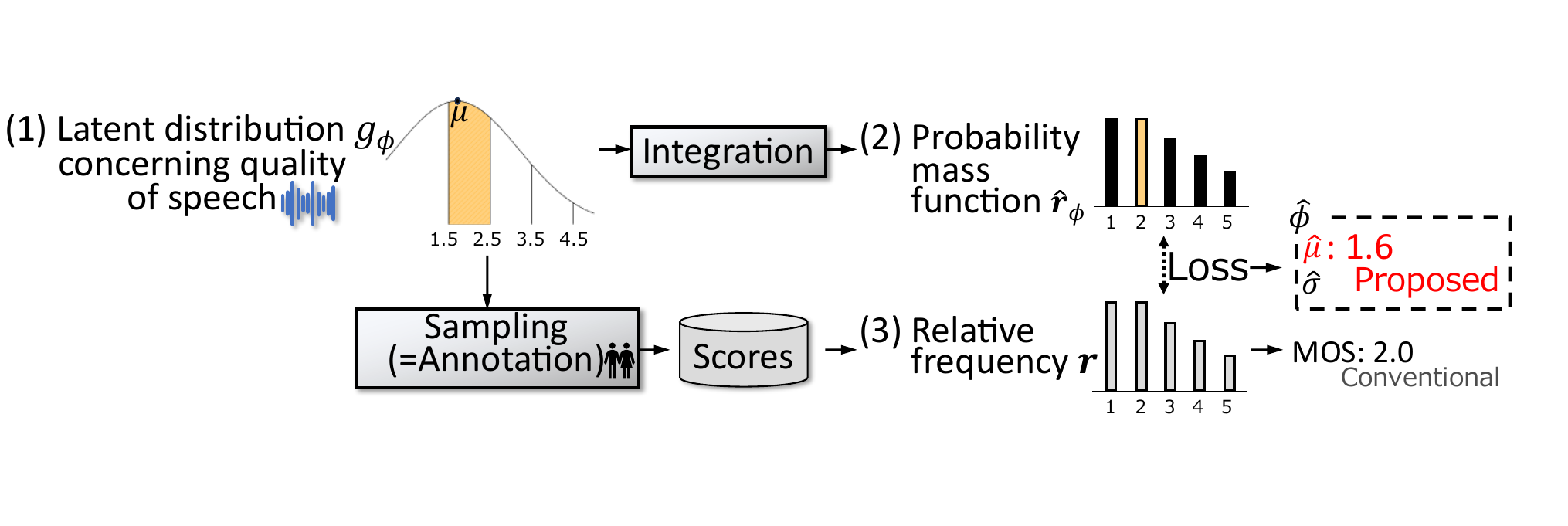}
\vspace*{-1.2cm}
 \caption{Overview of the proposed method. Continuous scores are assigned in annotators' mind, which form (1) an latent normal distribution. The proposed method estimates parameters of the latent distribution based on the loss between (2) the quantized distribution and (3) annotated rating distribution. We use the estimated peak $\hat{\mu}$ of the latent distribution as the new representative value instead of MOS.}
 \label{fig:top}
 \vspace*{-0.4cm}
\end{figure*}

While the superiority of $N_{\rm low}$-MOS over MOS has been experimentally confirmed, $N_{\rm low}$-MOS has three limitations: First, the appropriate use of $N_{\rm low}$-MOS is limited to the case where the number of ratings per audio sample is equal. Second, the appropriate value of $N$ for $N_{\rm low}$-MOS needs to be determined based on the number of ratings for each audio sample. Finally, since $N_{\rm low}$-MOS is based on the aforementioned hypothesis regarding {\it speech quality}, it may not be applicable to other subjective voice descriptors (SVDs)~\cite{kondo2024svd} such as cuteness and resonance. In this paper, we propose a new score aggregation method that does not suffer from these limitations.
An overview of the proposed method is shown in Figure \ref{fig:top}. The proposed method is based on an assumption different from that of $N_{\rm low}$-MOS, namely, that each annotator assigns a continuous score in their mind and selects the rating closest to the score from 1 to 5. By modeling this process, we model the generative distribution of ratings by quantizing the latent normal distribution. The proposed method estimates parameters of the latent distribution based on the loss between the quantized distribution and the annotated rating distribution, and uses the peak of the latent distribution as the new representative value instead of MOS. 
The proposed value is thought to be more appropriate than MOS through the above modeling since it tackles challenges (a) and (b) regarding ACR written in Sec.\ref{sec:motivation}.

The contributions of this study are as follows:
\begin{itemize}
\item We propose a new score aggregation method based on the modeling of the annotation process. This method is characterized by the fact that the representative value is calculated based on optimization.
\item We experimentally verify that replacing MOS with the new representative value improves comparison performance of MOSNet.
\end{itemize}

The structure of this paper is as follows: Section 2 reviews related work, Section 3 presents the proposed score aggregation method, Section 4 describes the experiments, and Section 5 provides the conclusion.
\section{Related work}
\subsection{SQA model}
\label{sec:sqa}
Since MOS tests for evaluation of TTS and VC systems require time-consuming questionnaires, neural-based MOS prediction models, such as MOSNet~\cite{lo2019mosnet}, have been developed. The input-output relationship of MOSNet can be expressed as $\hat{y}=f_{\theta}(\mathbf{X})$, where $\theta$ is the set of all network parameters to be trained, and $\mathbf{X}$ is the short-time Fourier transform (STFT) magnitude spectrogram of a speech sample. First, $\mathbf{X}$ is fed into the 2D convolutional layers to obtain the intermediate representation that captures the local time-frequency information. It is then input into bidirectional long short-term memories (BLSTMs) and two fully connected (FC) layers to obtain the frame-level scores $\hat{\bm{q}}=(\hat{q}_{1},\ldots,\hat{q}_{T})$, where $T$ represents the number of frames. Finally, the utterance-level score $\hat{y}$ is obtained by applying the average pooling to $\hat{\bm{q}}$.

The learning of $\theta$ is performed using the pair $(\mathbf{X}_{i}, y_{i})$, where $\mathbf{X}_{i}$ is the magnitude spectrogram of speech $i$, and $y_{i}$ is the MOS calculated as $y_{i}=1/N_{i}\sum_{n=1}^{N_{i}}s_{i,n}$, where $N_{i}$ is the number of socres annotated to speech $i$, and $\bm{s}_{i}=(s_{i,1},\ldots,s_{i,N_{i}})$ represents the $N_{i}$ ACRs annotated to speech $i$. The parameters are updated based on the backpropagation with the loss
\begin{align}
\label{eq:trainloss}
    \mathcal{L}_{\theta}=\left(\hat{y}_{i}-y_{i}\right)^2+\frac{\alpha}{T_i}\sum_{t}^{T_i}\left(\hat{q}_{i,t}-y_{i}\right)^2,
\end{align}
where $\hat{y}_{i}=f_{\theta}(\mathbf{X}_{i})$, $T_i$ is the number of frames of $\mathbf{X}_{i}$, $\hat{\bm{q}}_i=(\hat{q}_{i,1},\ldots,\hat{q}_{i,T_{i}})$ is the predicted frame-level MOS, and $\alpha$ is the weighting factor. The first term of $\mathcal{L}_{\theta}$ is the main term to make the output of $f_{\theta}$ closer to the actual MOS, while the second term is the auxiliary term to stabilize the predicted frame-level MOS.

Some approaches to enhance the performance of MOS prediction models have been proposed such as modification in model architecture \cite{shen2023non}, the use of self-supervised learning (SSL) models \cite{cooper2022generalization}, and the use of auxiliary information including listener IDs \cite{leng2021mbnet,huang2022ldnet}. UTMOS \cite{saeki2022utmos}, a prediction model that combines these approaches, achieved high performance in the VoiceMOS Challenge 2022 \cite{huang2022voicemos}, a competition for the SQA task after VC and TTS.
Additionally, exploring annotation question formats different from ACR and question schemes can also be considered as one approach~\cite{yasuda2023analysis}.

\subsection{Conventional score aggregation method \cite{kondo2024selecting}}
As an approach to improve the performance of MOS prediction, we have proposed a score aggregation method different from MOS, {\it $N_{\rm low}$-MOS} \cite{kondo2024selecting}. This method addresses the drawback that the reliability of MOS may be degraded by incorrect labeling, which can potentially restrict the performance of MOS prediction models. The attempt at a new score aggregation method is particularly relevant for training SQA models, as the number of ratings annotated for each audio sample is often limited due to the large number of audio samples required for training MOS prediction models. This is the opposite of typical MOS tests, where the impact of inappropriate annotations can be mitigated by averaging a large number of ratings. Unlike exploring annotation question formats as described in Sec.\ref{sec:sqa}, exploring score aggregation methods can be applied to prediction model training by recycling ratings, including SVD ratings, that were originally collected for purposes other than training prediction models.

For an audio sample $i$, let $s_{i, 1}, s_{i, 2}, ..., s_{i, N_{i}^{\rm (all)}}$ the $N_{i}^{\rm (all)}$ ACRs sorted in ascending order. The $N_{\rm low}$-MOS of speech sample $i$ is defined as $1/N \sum_{n=1}^{N} s_{i, n}$. Based on the hypothesis that the variance in ratings for each speech sample is primarily attributed to accidental inclusion of higher scores due to overlooking or neglecting the poor quality speech segments, $N_{\rm low}$-MOS is condisered to be more appropriate than regular MOS. We experimentally confirmed that by replacing the target of MOSNet from MOS to $N_{\rm low}$-MOS, the accuracy of quality comparison between speech samples improves~\cite{kondo2024selecting}. While the superiority of $N_{\rm low}$-MOS over MOS has been experimentally confirmed, $N_{\rm low}$-MOS has three limitations: First, the appropriate use of $N_{\rm low}$-MOS is limited to the case where the number of ratings per audio sample is equal. Second, the appropriate value of $N$ for $N_{\rm low}$-MOS needs to be determined based on the number of ratings for each audio sample. Finally, it may not be applicable to other subjective voice descriptors~\cite{kondo2024svd} since $N_{\rm low}$-MOS is based on the aforementioned hypothesis regarding speech quality.

Some SQA models have been proposed to predict not the representative value but the score distribution itself~\cite{patton2016automos}. The idea of score aggregation has the potential to be applied to the post-processing of methods that predict the distribution.
\section{Proposed method}
\subsection{Motivation}
\label{sec:motivation}
In this paper, we propose a new score aggregation method that can be applied even when the number of ratings varies for each audio sample, and does not require setting of $N$ in $N_{\rm low}$-MOS. For proposal, we focus on the following two limitations of ACR:

\begin{itemize}
\item[(a)] Regardless of how good or bad the audio quality is perceived, annotators are constrained to select from a given range of options, which can lead to the overestimation of low-quality speech or the underestimation of high-quality speech.
\item[(b)] Even when speech quality is perceived as intermediate between two consecutive choices, listeners are forced to either overestimate or underestimate the score they mentally assign. 
\end{itemize}
\subsection{Proposed score aggregation method}
The proposed method is based on the assumption that annotators assign continuous scores in their minds, and each annotator selects the rating closest to the score from 1 to 5. By modeling this annotation process, we model the generative distribution of ratings by quantizing the latent normal distribution as:
\begin{align}
g_{\phi}(u)&=\mathcal{N}(u;\mu,\sigma),\\
    \hat{\bm{r}}_\phi[1] &= \int_{-\infty}^{1.5} g_{\phi}(u) \, du, \label{eq:g1}\\
    \hat{\bm{r}}_\phi[k] &= \int_{k-0.5}^{k+0.5} g_{\phi}(u) \, du \quad (k=2,3,4), \\
    \hat{\bm{r}}_\phi[5] &= \int_{4.5}^{\infty} g_{\phi}(u) \, du,\label{eq:g5}
\end{align}
where $\mathcal{N}(u;\mu,\sigma)$ represents a latent normal distribution with mean $\mu$ and standard deviation $\sigma$, $\phi=\{\mu,\sigma\}$, and $\hat{\bm{r}}_\phi$ is the probability mass function of annotated ratings.

In the proposed aggregation method, the parameter $\phi$ of the above model is estimated, and the optimized $\mu$ is treated as the new representative value instead of MOS. The parameter $\phi$ is optimized by minimizing the following function:
\begin{align}
    \mathcal{L}_{\phi}=\mathcal{D}(\hat{\bm{r}}_\phi,\bm{r})+\beta(\sigma-\sigma_{\rm 0})^2, 
\end{align}
where $\bm{r}$ is the relative frequency distribution of the actual ratings, and $\mathcal{D}$ is an arbitrary divergence. The second term is a regularization term to prevent $\sigma$ from becoming too large when $\bm{r}$ is nearly uniform. $\beta$ is the weight coefficient, and $\sigma_{\rm 0}$ is the standard deviation directly calculated using the actual ratings. Although analytical minimization of $\mathcal{L}_{\phi}$ is difficult, we can obtain $\phi$ such that $\mathcal{L}_{\phi}<\mathcal{L}_{\phi_{\rm 0}}$ by applying an iterative minimization algorithm from the initial parameters $\phi_{\rm 0}=(\mu_{\rm 0},\sigma_{\rm 0})$, which consists of the mean $\mu_{\rm 0}$ (i.e., MOS) and $\sigma_{\rm 0}$ calculated using actual ratings. The proposed method tackles challenge (a) in Sec.~\ref{sec:motivation} due to the formulations of (\ref{eq:g1}) and (\ref{eq:g5}), and challenge (b) in Sec.~\ref{sec:motivation} by relaxed modeling using a latent continuous distribution and its integration.

Although a score aggregation technique that assumes a latent normal distribution has been previously applied to MOS research~\cite{kisida2024MOSshinri}, the research relies on an overly simplistic assumption that the variance of the score distribution is equal across all speech samples. This assumption contradicts the statistical report in \cite{lo2019mosnet}. In contrast, our approach achieves sophisticated aggregation by optimizing for each speech without imposing such strong assumption.
\section{Experiment}
In this chapter, we address the optimization problem for the proposed aggregation and experimentally verify that replacing MOS with the new representative value improves comparison performance of MOSNet.
\subsection{Experimental setting}
For representative values which are predicted by prediction models, we compared regular MOS~\cite{ITU-TP800}, $N_{\rm low}$-MOS~\cite{kondo2024selecting}, and the proposed value. The structure of the prediction models was the same as that of MOSNet~\cite{lo2019mosnet}, with only the predicted target being replaced by each representative value. The regularization weight $\alpha$ of model training was set to $\alpha=1$. For proposed method, L1 distance between cumulative distributions was used as $\mathcal{D}$:
\begin{align}
    \mathcal{D}(\hat{\bm{r}}_\phi,\bm{r})&=\sum_{k=1}^{5}\left|\hat{\bm{h}}_\phi[k]-\bm{h}[k]\right|,\\
\hat{\bm{h}}_\phi[k]&=\sum_{j=1}^{k}\hat{\bm{r}}_\phi[j],\\
\bm{h}[k]&=\sum_{j=1}^{k}\bm{r}[j].
\end{align}
This is a special case of the earth mover's distance~\cite{rubner1998metric}, the metric considering the distance between labels. For the regularization term, we set $\beta=0.03$. For optimization of $\mathcal{L}_{\phi}$, we used the iterative method SLSQP~\cite{kraft1988software} with a constraint $\sigma>10^{-5}$. The upper limit of the iteration number was set to 100. We used $\mu$ at the iteration where $\mathcal{L}_{\phi}$ was minimized as the representative value in the proposed method. For speech samples in which the proposed optimization did not yield ${\phi}$ such that $\mathcal{L}_{\phi}<\mathcal{L}_{\phi_{\rm 0}}$, then the regular MOS $\mu_{\rm 0}$ was used as the representative value of the proposed method.

For subjective speech quality dataset, we used the BVCC dataset~\cite{cooper2021voices}, which is composed of speech samples from various TTS and VC systems along with eight ACR scores for each sample. The numbers of speech samples in the training, validation, and test sets were 4974, 1066, and 1066, respectively. For $N_{\rm low}$-MOS, we used $N=6$, which showed the best performance in \cite{kondo2024selecting}.

For performance evaluation of prediction models, we used the utterance-level linear correlation coefficient (LCC)~\cite{pearson1920notes_MOSNet} and Spearman's rank correlation coefficient (SRCC)~\cite{spearman1961proof_MOSNet} between correct representative values, which are calculated using actual ratings, and predicted scores. LCC and SRCC are commonly used in conventional MOS prediction studies. While SRCC reflects the relative ranking among many speech samples, the reliability of the test ACRs used to calculate SRCC is limited by the difficulty of ACR annotation. Therefore, we also used ppref~\cite{carterette2008here}, the precision of the predicted scores to binary preference labels ($A\succ B$ and $A\prec B$): ${\rm ppref}=N_{\rm correct}/N_{\rm all}$, where $N_{\rm all}$ is the total number of binary labels, and $N_{\rm correct}$ is the number of the binary labels correctly classified by a prediction model. Since binary ranking labels are easier to annotate than ACR, ppref is expected to be a more accurate metric than SRCC. Additionally, while the results of ACR annotation are influenced by the overall sound quality of the given speech dataset~\cite{cooper2023investigating}, relative labels are thought to be less affected. We prepared 150 questions for ppref; Each question consists of two sequentially played speech samples (Speech A and Speech B) with 1.5 second gap from BVCC test dataset. Annotators selected one of four options `A (sure)', `A (not sure)', 
`B (not sure)' and `B (sure)', focusing speech quality. The pairs were limited to those with a MOS difference between 0.5 and 1. Additionally, among the 150 pairs, 50 included at least one speech sample with ${\rm MOS} \leq 2$, another 50 included at least one sample with ${\rm MOS} \geq 4$, and the remaining pairs consisted solely of samples with $2 < {\rm MOS} < 4$. We screened a subset using the following procedure: Four annotators answered all questions. For calculating ppref, we retained only questions where at least three annotators agreed on 'A (sure)' or 'A (not sure)' with none selecting 'B (sure)' as $A\succ B$, and vice versa for $A\prec B$. Finally, the total number of binary rankings used for ppref was $N_{\rm all}=106$. Even with this limited number, the test dataset still holds significant value for measuring comparison ability, as it consists of speech sample pairs with varying sound quality, each constructed from samples with similar MOS.

For speech data, the sampling frequency was set to 16 kHz. For short-time Fourier transform, a Hamming window was used with a window length and shift width of 32 ms and 16 ms, respectively. Adam optimizer was used for training with a learning rate of $10^{-4}$. The batch size was set to 32. Early stopping based on SRCC calculated on the validation set was employed with a patience of 15. The dropout rate was set to 0.3. Each initial parameter was initialized with a standard normal distribution, and the number of initialization trials was set to 8.

\subsection{Optimization result}
\begin{figure}[t]
\vspace*{0cm}
 \centering
 \hspace*{-0.0cm}
\includegraphics[width=1.0\columnwidth]{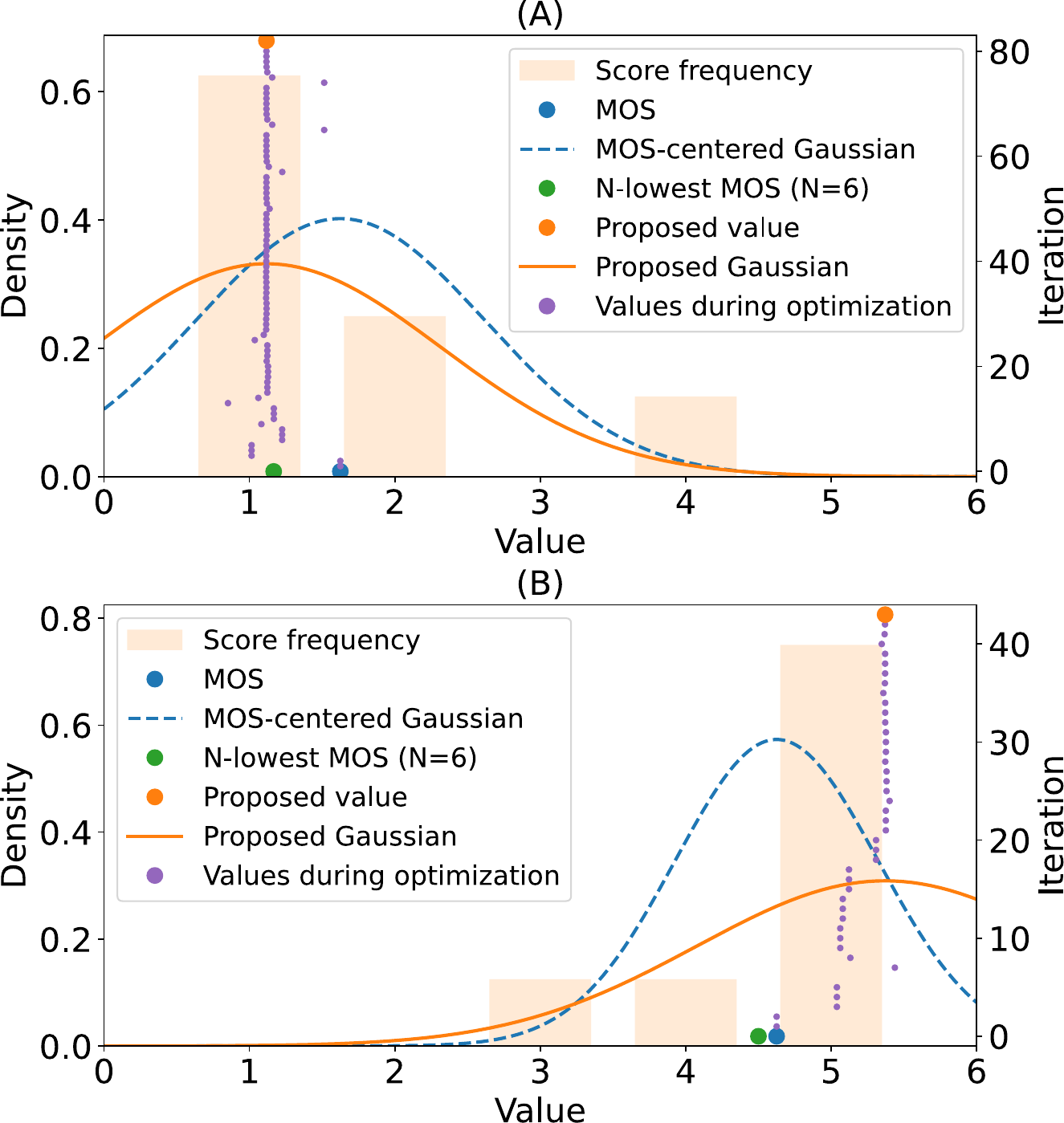}
\vspace*{-0.0cm}
 \caption{
Examples of rating histograms, MOS, $N_{\rm low}$-MOS, and the optimization process of the proposed method for a single speech sample. (A) shows an example for a speech sample with a low MOS, and (B) shows an example for a speech sample with a high MOS.}
 \label{fig:example}
 \vspace*{-0.0cm}
\end{figure}
\begin{figure}[t]
\vspace*{0cm}
 \centering
 \hspace*{-0.0cm}
\includegraphics[width=1.0\columnwidth]{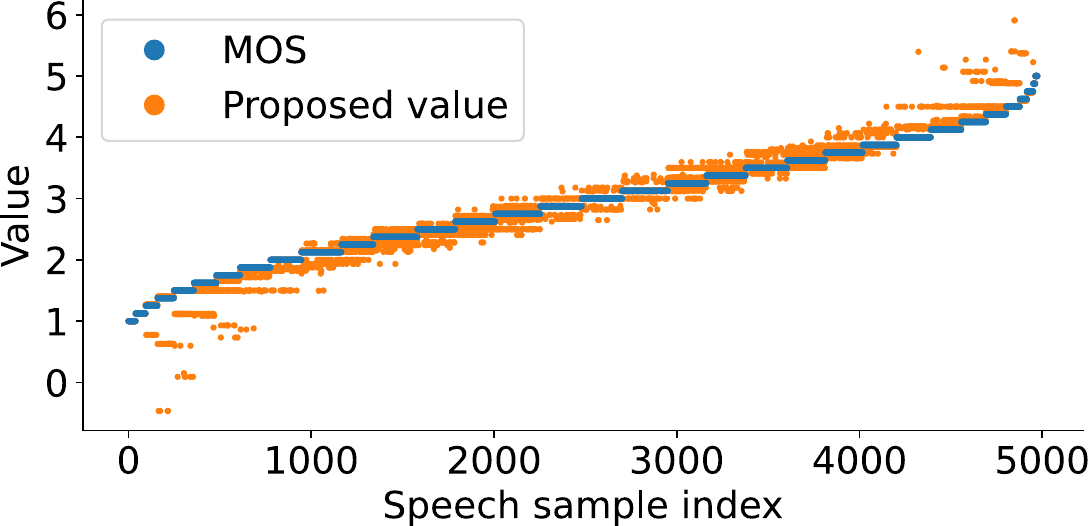}
\vspace*{-0.0cm}
 \caption{MOS and proposed representative value of speech quality for each speech sample.}
 \label{fig:all}
 \vspace*{-0.0cm}
\end{figure}
\begin{table}[t]
\caption{The average of 8 trials for prediction performance.}
\label{tab:result_ave}
\tabcolsep = 1.5pt
\centering
{
 \begin{tabular}{c|c|c|c} \bhline{1.5pt}
 \begin{tabular}{c}\end{tabular}&\begin{tabular}{c}
LCC
\end{tabular} &\begin{tabular}{c}
SRCC
\end{tabular}&\begin{tabular}{c}
Ppref
\end{tabular} \\ \bhline{1.5pt}
\begin{tabular}{c}
MOS
\end{tabular}&0.638$\pm$0.028&0.632$\pm$0.031&\begin{tabular}{c}
0.716\\$\pm$0.039\end{tabular}\\\hline
\begin{tabular}{c}
$N_{\rm low}$-MOS\\ ($N=6$)
\end{tabular}&\begin{tabular}{c}
\textbf{0.661}$\pm$\textbf{0.046}\end{tabular}&\begin{tabular}{c}
\textbf{0.656}$\pm$\textbf{0.048}\end{tabular}&\begin{tabular}{c}
0.730\\$\pm$0.029\end{tabular}\\\hline
\begin{tabular}{c}
Proposed value
\end{tabular}&\begin{tabular}{c}
0.651$\pm$0.040\end{tabular}&\begin{tabular}{c}0.641$\pm$0.044
\end{tabular}&\begin{tabular}{c}\textbf{0.737}\\$\pm$\textbf{0.019}
\end{tabular}\\
\bhline{1.5pt}
 \end{tabular}
}
\vspace*{-0.4cm}
\end{table}

\begin{table}[t]
\caption{Prediction performance in the trial where SRCC is maximized on the validation dataset.}
\label{tab:result_max}
\tabcolsep = 1.5pt
\centering
{
 \begin{tabular}{c|c|c|c} \bhline{1.5pt}
 \begin{tabular}{c}\end{tabular}&\begin{tabular}{c}
LCC
\end{tabular} &\begin{tabular}{c}
SRCC
\end{tabular}&\begin{tabular}{c}
Ppref
\end{tabular} \\ \bhline{1.5pt}
\begin{tabular}{c}
MOS
\end{tabular}&0.667&0.681&0.708\\\hline
\begin{tabular}{c}
$N_{\rm low}$-MOS ($N=6$)
\end{tabular}&\begin{tabular}{c}
\textbf{0.722}\end{tabular}&\begin{tabular}{c}
\textbf{0.724}\end{tabular}&\begin{tabular}{c}
\textbf{0.764}\end{tabular}\\\hline
\begin{tabular}{c}
Proposed value
\end{tabular}&\begin{tabular}{c}
0.702\end{tabular}&\begin{tabular}{c}0.693
\end{tabular}&\begin{tabular}{c}0.726
\end{tabular}\\
\bhline{1.5pt}
 \end{tabular}
}
\vspace*{-0.4cm}
\end{table}
Among all 7106 speech samples, 6972 yielded a ${\phi}$ such that $\mathcal{L}{\phi}<\mathcal{L}{\phi_{\rm 0}}$ through optimization. For the 85 samples where all the ratings were the same, a $\phi$ such that $\mathcal{L}{\phi}<\mathcal{L}{\phi_{\rm 0}}$ could not be obtained, and as a result, the representative value was the rating itself. Figure \ref{fig:example} shows the frequency distribution of annotated ratings, MOS, $N_{\rm low}$-MOS, and the optimization process of the proposed method for two speech examples (one with a low MOS and another with a high MOS). It also illustrates the latent normal distributions before and after optimization. It can be confirmed that the optimized latent distribution more accurately captures the shape of the frequency distribution in both examples, resulting in the proposed value close to $N_{\rm low}$-MOS for the low MOS example but far from $N_{\rm low}$-MOS for the high MOS example. During the optimization process, several iterations where the loss significantly increased were observed. This corresponds to the rise in the representative value between the 60th and 80th iterations shown in Fig.~\ref{fig
} (A). This is attributed to the complexity of the function being optimized by SLSQP. Exploring more suitable optimization methods is a future work.

Figure \ref{fig:all} displays MOS and proposed values of each training sample sorted in ascending order corresponding to MOS. The proposed values can take on values outside the range from 1 to 5, as also seen in Fig.~\ref{fig:example} (B). This is because the proposed value is influenced by the detailed information of the frequency distribution shape, including variance and skewness. Additionally, the proposed representative values tend to deviate further from the central score of 3 compared to the MOS. This indicates that the proposed representative values suppress the overestimation of low MOS samples and the underestimation of high MOS samples.

\subsection{Prediction result}
Table \ref{tab:result_ave} presents prediction performance averaged across all trials. Additionally, since the best-performing model can be selected from several trials for actual use cases, we show the prediction performance of the trial that yielded the highest SRCC on the validation data in Tab.~\ref{tab:result_max}. In both tables, the proposed method achieves higher LCC, SRCC, and ppref than MOS. This performance improvement is attributed to the proposed method addressing challenges (a) and (b) written in Sec.\ref{sec:motivation} through more appropriate formulations.

Although the proposed method underperforms $N_{\rm low}$-MOS in all metrics except for the average ppref, the proposed method can be applied even when the number of ratings varies for each audio sample, and does not require setting of $N$ in $N_{\rm low}$-MOS.

While the SRCC in Tab.~\ref{tab:result_max} significantly exceeds the averaged SRCC shown in Tab.~\ref{tab:result_ave}, the ppref in Tab.~\ref{tab:result_max} is lower than the averaged ppref in Tab.~\ref{tab:result_ave}. This is because the number of pairs used to calculate ppref was very limited in this study, resulting in ppref not being representative of the entire set of BVCC speech samples. This is contrast to SRCC, which is calculated using a very large number of speech samples. However, considering the difficulty of ACR annotation, we recommend that colleagues use not only ACR-based metrics but also ppref for performance evaluation.

\section{Conclusion}
We introduced a novel score aggregation method to address the limitations of conventional annotations for MOS. This method is grounded in the hypothesis that annotators mentally assign continuous scores and subsequently select the closest discrete rating. By quantizing the latent continuous distribution, we approximated the generative distribution of ratings and used the peak of this latent distribution as a new representative value in place of MOS. Experimental results demonstrated that replacing MOSNet's predicted target with this proposed value enhances LCC, SRCC, and ppref. This performance improvement is attributed to the proposed method's effective handling of several challenges inherent in ACR annotation. Future work will explore the development of more suitable optimization algorithms for the proposed formulation. This work can motivate researchers interested in SQA to further explore aggregation methods to develop models that more accurately quantify subjective assessments of speech.


\bibliographystyle{IEEEbib}
\bibliography{refs}

\end{document}